\documentclass{pos}
\usepackage{epsfig}
\newcommand{\pathnow}{}
\title{Particle Production and Deconfinement Threshold }

\ShortTitle{Particle Production and Deconfinement Threshold}

\author{\speaker{Johann Rafelski}%
\\
         University of Arizona, Tucson, AZ 85721, USA and\\ Department f\"ur Physik der
Ludwig-Maximilians-Universit\"at M\"unchen,\\
Maier-Leibnitz-Laboratorium, Am Coulombwall 1, 85748 Garching, Germany\\
         E-mail: \email{Rafelski@Physics.Arizona.EDU}}

\author{Jean Letessier\\
LPTHE, Universit\'e Paris 7, 2 place Jussieu, F--75251 Cedex 05}        
\abstract{
We present a detailed analysis of the NA49 experimental particle yield 
results,  and 
discuss the physical properties of the particle source. We explain in depth  
how our analysis differs from the work of other groups, what advance this 
implies in terms of our understanding, and what
new physics about the deconfined particle source this allows us to recognize.
We answer several frequently asked questions, presenting a transcript of a discussion 
regarding our data analysis. We show that the 
final NA49 data at 40, 80, 158 $A$GeV lead to a remarkably constant
extensive thermal chemical-freeze-out  properties of the fireball. We discuss briefly 
the importance of thermal hadronization pressure.
}

\FullConference{8th Conference Quark Confinement and the Hadron Spectrum\\
                 September 1-6, 2008\\
                 Mainz. Germany}

\begin{document}
\section{Deconfinement and Hadron Production}
\begin{figure}[htb] 
\hspace*{0.7cm}\epsfig{width=8cm, angle=-90,figure=\pathnow 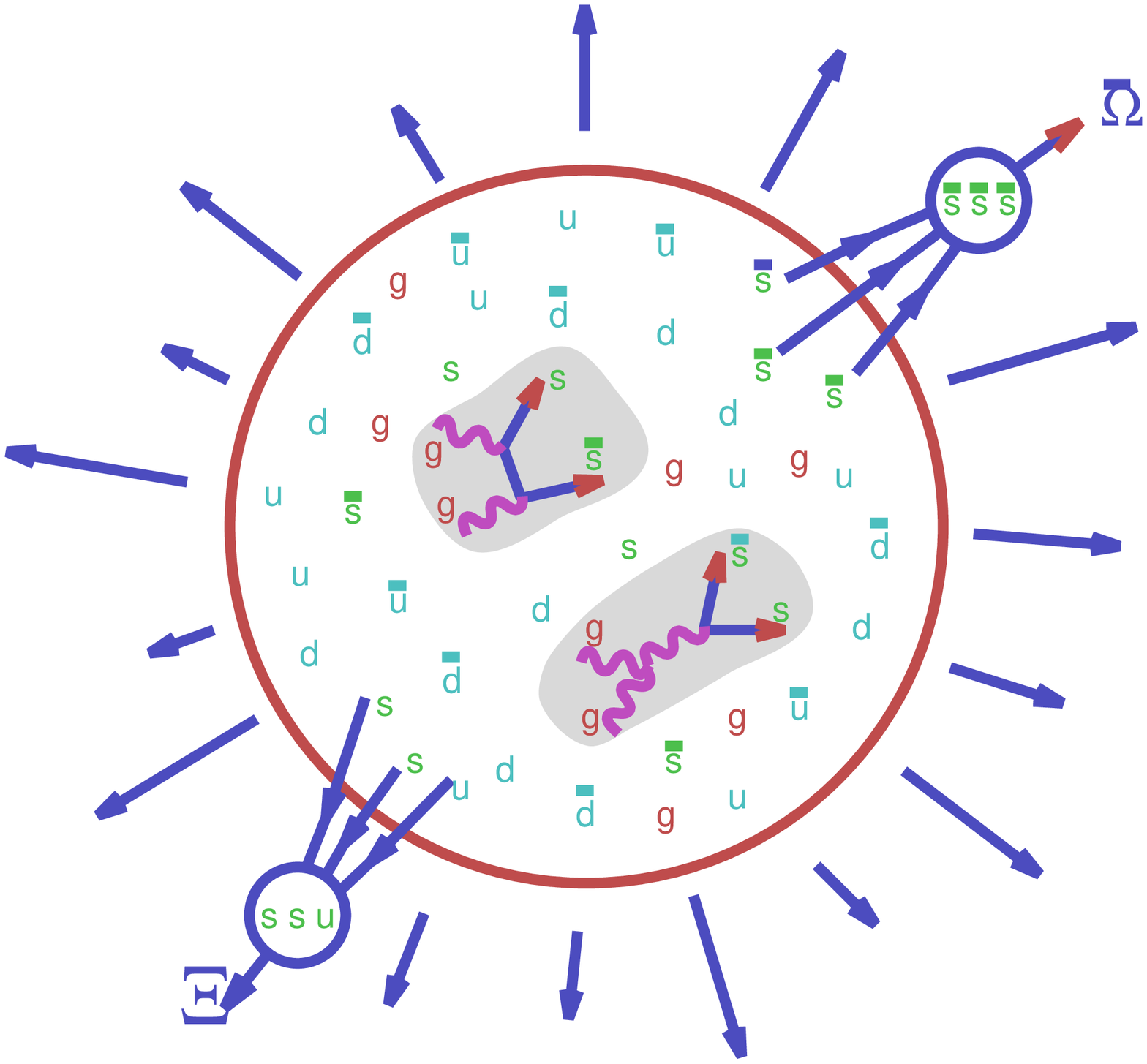}\vskip -8cm\hspace*{8cm}
\epsfig{width=7cm,figure=\pathnow 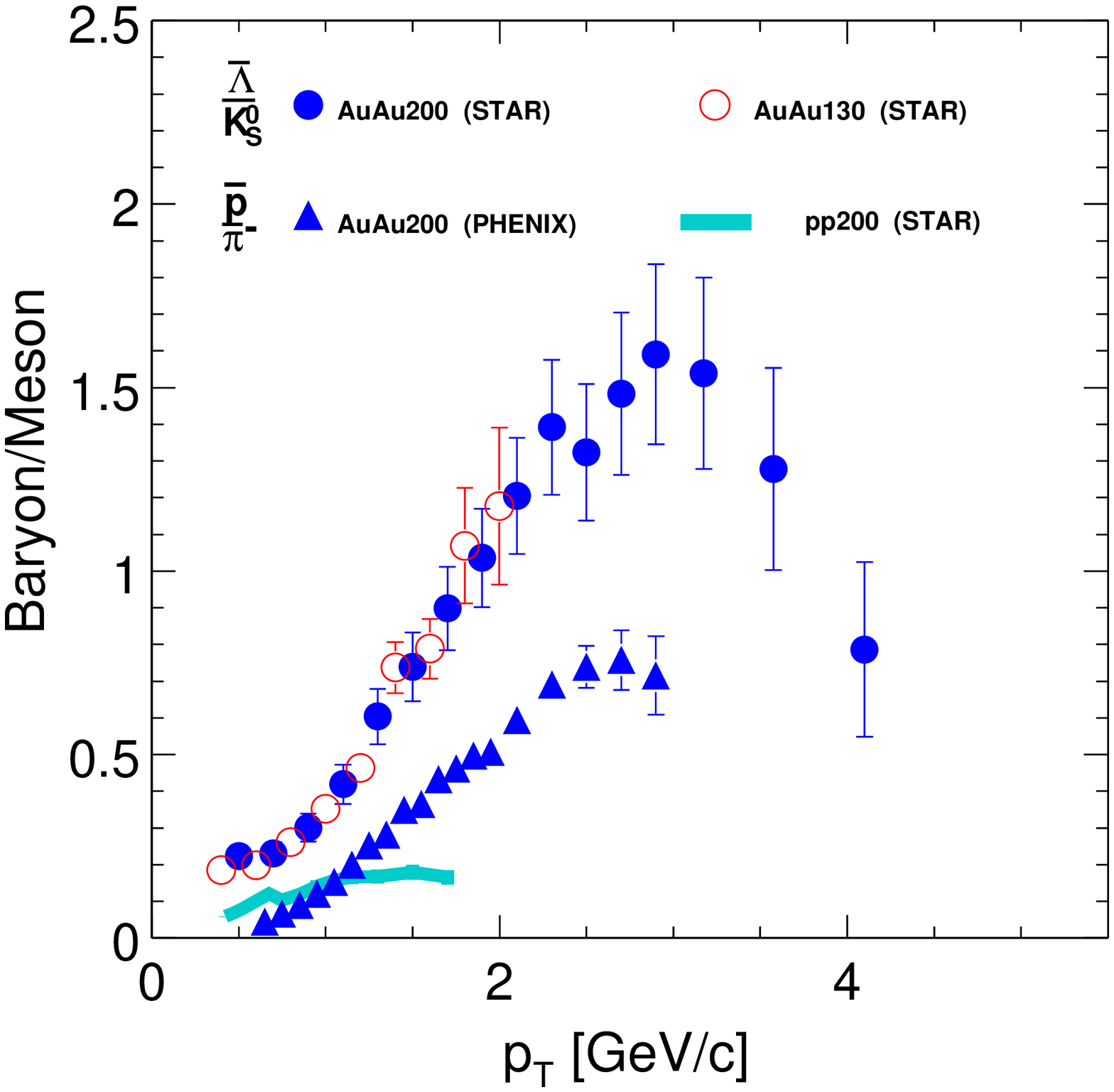} 
\caption{\label{JRSTRANGEPRODa}%\caption{\label{JRlkppi}
On left, qualitative  illustration of (strange) hadron production in QGP: 
gluon fusion reactions populate flavor yield, subsequently at phase breakup, quarks merge into 
final state hadrons. 
On right, example of  baryon to meson ratio 
experimental results  ${\overline{\Lambda}}/K_{S}$ and $\overline{p}/\pi$ 
in  Au-Au compared to $pp$ collisions  as a function of  $p_\bot$. 
}
\end{figure}

Enhanced production of strange hadrons, and of (strange) antibaryons is
a signature of quark--gluon plasma (QGP) formation  in  relativistic
heavy ion  collisions~\cite{Rafelski:1980rk}. We illustrate on left 
in figure \ref{JRSTRANGEPRODa} the two step  mechanism 
of hadron production from quark--gluon matter: upon deconfinement thermal glue emerges
from parton matter, thermal gluon fusion reactions produce  quark pairs 
in mass range $m_i<3T$~\cite{Muller:1983ed}. Subsequently, at
a later time and thus at a  much lower ambient temperature,
quarks merge into final state hadrons~\cite{Koch:1986ud,BMCombine}. 
On the right in figure  \ref{JRSTRANGEPRODa} we show the relative yield as function of $p_\bot$ 
of baryon to meson abundance
which are quite different from what is found  in  $pp$ reactions~\cite{Huang:2005nda}. 
This confirms that the relativistic heavy ion  hadron production differs from 
the string breaking mechanism of hadron production~\cite{Andersson:1992iq}.
This result found also at top SPS energy range~\cite{Laszlo:2008gf}.

%%%%%%%%%%%%%%%%%%%%%%%%%%% fig 1
%%%%%%%%%%%%%%%%%%%%%%%%%%%

Of particular relevance in this work is the
parameter controlling the absolute yield of quark pairs $\gamma_q^{\rm QGP}$  for  
$u,\ d$ and  $\gamma_s^{\rm QGP}$ for $s$ (and $\gamma_c^{\rm QGP},\ \gamma_b^{\rm QGP}$ for $c,\ b$
not further discussed here). In an equilibrating system all $\gamma_i^{\rm QGP}\to 1$  
as function of time, one can argue that at RHIC strange flavor in QGP is nearly 
equilibrated. However, in an experiment, we observe 
hadrons, these yields are subject to different parameters operating 
in the hadron phase (HP) 
with the initial values $\gamma_i^{\rm HP}(t_f)$ If hadrons do not scatter much after
formation,  as is the case in the single freeze-out model~\cite{Broniowski:2008vp,Torrieri:2000xi},
there is little if any change of $\gamma_i^{\rm HP}(t_f)$.

Remarkably, these two particle yield parameters, 
 $\gamma_q^{\rm HP}(t_f)$ and $\gamma_s^{\rm HP}(t_f)$,  also  control
 the relative yield of baryons to mesons
shown on right  in figure \ref{JRSTRANGEPRODa},
  $$ {{\rm baryons}\over {\rm mesons}}\propto 
{\gamma_q^{{\rm HP}\,3}\over \gamma_q^{{\rm HP}\,2}}
\cdot \left({\gamma_s^{\rm HP}\over \gamma_q^{\rm HP}}\right)^n \propto \gamma_q^{\rm HP},$$  
where  $ \gamma_s/\gamma_q$ shifts the yield of strange  vs non-strange hadrons, thus for example
the relative yields of hadrons obey:
$$
{{\overline {\Lambda}(\bar u\bar d\bar s)} \over 
{\bar p(\bar u\bar u\bar d)}}\propto{ \gamma_s^{\rm HP}\over\gamma_q^{\rm HP}},
\qquad
{{\rm K}^+(u\bar s)\over \pi^+(u\bar d)}\propto{ \gamma_s^{\rm HP}\over\gamma_q^{\rm HP}} \,,
\qquad 
{\phi\over h}\propto { \gamma_s^{2\,{\rm HP}}\over\gamma_q^{2\,{\rm HP}}}\,,
\qquad 
{\Omega(sss)\over \Lambda(sud)}\propto { \gamma_s^{2\,{\rm HP}}\over\gamma_q^{2\,{\rm HP}}}\,.
$$

The above observation, when combined with the results shown on right 
in figure~\ref{JRSTRANGEPRODa},  imply
that the baryon to meson yield and thus  $\gamma_q^{\rm HP}(t_f)$
is arising from  microscopic dynamics of hadronization. This establishes 
the necessity to include the occupancy parameters in 
 order to describe the yields of hadrons, since this is the parameter which 
allows   for a reaction dependent relative yield of mesons and baryons. 
Conversely,  a study of particle yields with a 
fixed  value $\gamma_q^{\rm HP}=1$   
presumes certain relative yield of baryons to mesons 
and this is  over-constraining a model of hadronization. Therefore, in such
an approach, in order to describe the data precisely, the control of
the relative number of produced mesons and baryon is achieved by manipulating  
the hadron spectrum~\cite{PBMretraction} --- 
including  additional meson and baryon resonances,
it is possible to fine tune the final relative yield of meson and baryons. Naturally,
such a fine tuning of the hadron mass spectrum in fact confirms the failure
of the   `equilibrium' model~\cite{PBMequilibrium}, and confirms the necessity 
to use the non-equilibrium occupancy parameters, if not for any fundamental reason,
than in order to fudge the errors in the mass spectrum of hadrons.

The greatest experimental sensitivity of the baryon to meson ratio is present
when the respective yields are comparable,  such that the particle yields
are a sensitive measure of this ratio. This is the 
case for the intermediate energy range at 
SPS, e.g., in the 80 $A$GeV Pb collisions 
with laboratory fixed target. Our choice to consider 
the SPS energy range in this presentation  is in 
part dictated by this observation, and in part by
the fact that we expect a change in reaction mechanism 
at an energy range below the 
RHIC `high' energy runs. 

Other statistical hadronization (SH) parameters we derive from  the data are
the source volume $V$ and the temperature  $T$, at which 
particles stop changing in yield (chemical freeze-out).  Moreover, we obtain
chemical potentials $\mu_B=3\mu_q=3T\ln \lambda_q,\ \mu_S=  T\ln (\lambda_q/\lambda_s)$,
related to conserved quantum numbers:
baryon number and  strangeness, respectively. We also obtain $\lambda_{I3}$
which allows to conserve the net charge and expresses 
the asymmetry in the 3-rd component of the isospin. Especially for low 
energy reactions where the particle yield is relatively low this parameter differs 
significantly from unity. 
 
The particle yields measure the values of SH parameters assuming that all particles 
stop changing their abundance nearly at the same point in time. Here, the yield of
stable hadronic particles (ignoring weak decays) includes hadronic resonance
decay chain. Note that even if hadronic resonances continue to react and their 
number changes in time after QGP breakup,  
the final number of stable particles is unchanged. 
Thus, the chemical freeze-out values of SH parameters, and the associated physical
properties of the system at hadronization can be related to the QGP breakup 
condition, if these hadrons are free-streaming out of the deconfined
fireball. 

Since the phase space density is in general different in the two phases, in order
to preserve entropy (the valance quark pair number)
across the phase boundary there must be  a jump in the phase
 space occupancy parameters $\gamma_q$ ---
 this effect replaces the increase in volume in a 
slow re-equilibration with  mixed phase which
accommodates transformation of a QGP entropy 
dense phase into HP dilute phase. Similarly, 
there is a jump in strangeness occupancy since 
QGP is a more strangeness dense phase than is HP.
$\gamma_i^{\rm HP}(t_f)  > \gamma_i^{\rm QGP}(t_f),\ i=q,\ s$, again
due to rather large hadron masses. In order to relate  
the initial values $\gamma_i^{\rm HP}(t_f),\ i=q,\ s$, at freeze-out (subscript-$f$),
to the QGP fireball source value $\gamma_i^{\rm QGP}(t_f),\ i=q,\ s$, we
match across phase boundary by considering the continuity of entropy  
and strangeness. In principle, only the 
occupancies are discontinuous at QGP hadronization, the other SH
parameters are  smooth.  For this reason,  we   attached 
phase-indices `QGP' and `HP' only to occupancy parameters, and when these
are omitted we always mean implicitly the confined hadron phase HP.

%%%%%%%%%%%%%%%%%%%%%%%%%%%%%%%%%%%%%%%%%%%%%%%%%%%%%%%%%%%%%%%
\section{Hadron Yields: Data and Fits}

A complete analysis of experimental hadron yield results 
requires a significant book-keeping and fitting  effort in 
order to allow for resonances, particle
widths, full decay trees, isospin multiplet sub-states.
A program SHARE (Statistical HAdronization with REsonances)
suitable to perform this data analysis 
is available for public use~\cite{Torrieri:2004zz,Torrieri:2006xi}.
This program implements the PDG~\cite{Yao:2006px} confirmed (4-star) 
set of particles and resonances, and we use~\cite{Letessier:2005qe}
the  recent determination  of $\sigma$-meson mass~\cite{Yndurain:2007qm}
($m_\sigma=484,\ \Gamma_\sigma/2=255$ MeV).
The data set we use in table \ref{AGSSPS} is the latest NA49-2008 results~\cite{NA49},
together  with, as reference, AGS data set at their highest energy,
 see~\cite{Letessier:2004cs,Letessier:2005qe}.  A star at $\lambda_s$
indicates that we  fixed the value by   strangeness conservation, which we do at AGS  
considering a small data set.

%%%%%%%%%%%%%%%%%%%%%%%
\begin{table*}[!t]
\caption{ 
AGS (on left)  and SPS energy range particle multiplicity data sets used in fits (see text). 
In bottom of table, we show  the fitted statistical parameters and the corresponding 
chemical potentials. 
\label{AGSSPS}}
%\small
\scriptsize
\begin{center}
\begin{tabular}{|l| c | c c c c c |  }
\hline
E[$A$GeV]                      & 11.6          & 20          & 30              & 40         & 80            & 158  \\
$\sqrt{s_{\rm NN}}$  [GeV]     &4.84           &6.26         &7.61             &8.76          &12.32          &17.27  \\
$y_{\rm CM}$                   &1.6  &1.88 &2.08 &2.22 & 2.57& 2.91 \\
%\hline
$N_{4\pi}$ centrality          &most central   &  7\%          &  7\%          &  7\%           &  7\%           &  5\%\\
\hline
 $N_W $, AGS: $p/\pi^+$
                        &$1.23 \pm 0.13$  &349$\pm$6      &349$\pm$6      &349$\pm$6      &349$\pm$6      &362$\pm$6           \\
$Q/b $                      &0.39$\pm$0.02  &0.394$\pm$0.02 &0.394$\pm$0.02 &0.394$\pm$0.02 &0.394$\pm$0.02 &0.39$\pm$0.02   \\
$(s-\bar s)/(s+\bar s)$    & $0\pm0.05$    & $0\pm0.05$ & $0\pm0.05$ &$ 0\pm0.05$ & $0\pm0.05$   & $0\pm0.05$   \\
 \hline
$\pi^+$                     &133.7$\pm$9.9  &190.0$\pm$10.0 &241$\pm$13   &293$\pm$18     &446$\pm$27     &619$\pm$48       \\
$\pi^-$, AGS: $\pi^-\!/\pi^+$ 
                        & $1.23 \pm 0.07$ &221.0$\pm$12.0   &274$\pm$15 &322$\pm$19     & 474$\pm$28     &639$\pm$48       \\
${\rm K}^+$, AGS: ${\rm K}^+\!/{\rm K}^-$ 
                            &$5.23\pm0.5$  &40.7$\pm$2.9     &52.9$\pm$4.2   &56.1$\pm$4.9  &73.4$\pm$6     &103$\pm$10      \\
${\rm K}^-$                 &3.76$\pm$0.47  &10.3$\pm$0.3  &16$\pm$0.6     &19.2$\pm$1.5    &32.4$\pm$2.2   &51.9$\pm$4.9      \\
$\phi $, AGS: $\phi/{\rm K}^+$ 
                         &$0.025\pm 0.006$&1.89$\pm$0.53  &1.84$\pm$0.51   &2.55$\pm$0.36   &4.04$\pm$0.5   & 8.46$\pm$0.71      \\
$\Lambda$                   &18.1$\pm$1.9   &27.1$\pm$2.4     &36.9$\pm$3.6   &43.1$\pm$4.7   &50.1$\pm$  10    &44.9$\pm$8.9       \\
$\overline\Lambda$          &0.017$\pm$0.005&0.16$\pm$0.05  &0.39$\pm$0.06  &0.68$\pm$0.1   &1.82$\pm$0.36  &3.68$\pm$0.55    \\
$\Xi^-$                     &               & 1.5$\pm$0.3  & 2.42$\pm$0.48 &2.96$\pm$0.56  &3.8$\pm$0.87  & 4.5$\pm$0.20      \\
$\overline\Xi^+$            &               &               &0.12$\pm$0.05  & 0.13$\pm$0.03 &0.58 $\pm$0.19 & 0.83$\pm$0.04      \\
%$\Omega$                   &               &               &               &                &              &                     \\
%$\overline\Omega$          &               &               &               &                &              &                   \\
$\Omega+\overline\Omega$, or    &               &               &               &0.14$\pm$0.07   &              &                   \\[-0.25cm]
\hspace*{1.2cm}${\rm K}_{\rm S}$           &               &               &               &                &              & 81$\pm$4           \\
%${\rm K}(892)0  $          &               &               &               &                &              &                   \\
%${\rm K}(892)0/{\rm K}^-$  &               &               &               &                &              &                   \\
 \hline 
$V {\rm [fm}^3]$            &3649$\pm$331   &4775$\pm$261   &2229$\pm$340   &1595$\pm$383    &2135$\pm$235   & 3055$\pm $454  \\
$T$ [MeV]                   &153.5$\pm$0.8  &151.7$\pm$2.8  &123.8$\pm$3    &130.9$\pm$4.4   &135.2$\pm$0.01 &136.0$\pm$0.01       \\
$\lambda_q^{\rm HP}$                 &5.21$\pm$0.07  &3.53$\pm$0.09  &2.86$\pm$0.09  &2.42$\pm$0.09   &1.98$\pm$0.07 & 1.744$\pm$0.02     \\
$\lambda_s^{\rm HP}$                 &1.565$^*$      &1.39$\pm$0.05  &1.45$\pm$0.05  &1.34$\pm$0.06   &1.25$\pm$0.18  & 1.155$\pm$0.03   \\
$\gamma_q^{\rm HP}$                 &0.366$\pm$0.008  &0.49$\pm$0.03  &1.54$\pm$0.37  &1.66$\pm$0.14  &1.65$\pm$0.01 &1.64$\pm$0.01\\
$\gamma_s^{\rm HP}$                  &0.216$\pm$0.009 &0.40$\pm$0.03  &1.61$\pm$0.07  &1.62$\pm$0.25  &1.52$\pm$0.06 & 1.63$\pm$0.02     \\
$\lambda_{I3}^{\rm HP}$             &0.875$\pm$0.166  &0.877$\pm$0.05 &0.935$\pm$0.013&0.960$\pm$0.027&0.973$\pm$0.014& 0.975$\pm$0.005     \\
\hline
$\mu_{\rm B}$ [MeV]               &759      &574            &390            &347             &276           & 227   \\
$\mu_{\rm S}$ [MeV]               &180      &141            & 83.7          &77.6            &62.0          & 56.0  \\
 \hline
\end{tabular}
\end{center}
 \end{table*}
%%%%%%%%%%%%%%%%%%%%%%%

Aside of directly measured yields, we  fit the baryon content (`measured' in terms of
event  centrality choice), the  charge per baryon (`independently' measured  to 
be the proton content in nuclei) and the 
strangeness balance, all three shown in separate top data section of the table. 
Since  `strangeness content' $s+\bar s$ is large,
we choose to consider  $\delta s=(s-\bar s)/(s+\bar s)$.
The errors we present for these entries are our estimates of how the error in measurement 
propagates into the statistical parameters.

%%%%%%%%%%%%%%%%%%%%%%%%%%%%%%%%%%%%%%%%%%%%%%%%%%%%%%%%%%%%%%%%%%%%%%%%%%%%%%%%%%%%%

\begin{table}[t]
\caption{
\label{outputAGSPS}Predictions: AGS/SPS particle yields   including NA49 2008 data, all SHARE 2.2.  
}
\vspace*{0.2cm}
\small   
%\scriptsize
\begin{center}
\begin{tabular}{|c| c | c c c c c | }
\hline
$E$ [$A$\,GeV]             &11.6 & 20  & 30    & 40    & 80  & 158  \\ 
$\sqrt{s_{\rm NN}}$  [GeV] &4.84 &6.26 &7.61   &8.76   &12.32 &17.27  \\ 
$y_{\rm CM}$               &1.6  &1.88 &2.08   &2.22   & 2.57& 2.91 \\
\hline
$N_{4\pi}$/ centr.      &m.c. &  7\% &  7\% &  7\% &  7\% &  5\%\\
\hline
$b\equiv B-\overline B$    & 375.6&348.1 &348.6 &349.9 &349.5 &361.7 \\ 
$(s-\bar s)/(s+\bar s)$    & 0    &-0.119&-0.037 &-0.007&-0.017&-0.064  \\
\hline
$\pi^+$                    & 134.0&189.9 &243.3 &292.5 &434.7 &617.2 \\ 
$\pi^-$                    & 161.2&223.4 &278.5 &324.1 &469.6 &663.7 \\ 
${\rm K}^+$                & 17.5 &41.1  &50.2  &53.4  &72.7  &111.3 \\ 
${\rm K}^-$                & 3.60 &10.3  &15.9  &19.7  &33.4  &54.7  \\ 
${\rm K}_{\rm S}$          & 10.9 &26.3  &33.1  &36.2  &52.1  &81.3  \\ 
$\phi $                    & 0.47 &1.82  &2.10  &2.64  &4.23  &7.37  \\ 
$p $                       &173.2 &162.9 &166.2 &137.2 &138.6 &145.9 \\ 
$\bar p$                   &0.022 &0.207 &0.57  &0.74  &2.46  &5.39  \\ 
$\Lambda$                  & 18.7 &29.3  &39.5  &36.9  &41.3  &48.5  \\ 
$\overline\Lambda$         &0.016 &0.16  &0.40  &0.62  &1.77  &4.02  \\ 
$\Xi^-$                    & 0.49 &1.34  & 2.45 &2.77  &3.42  &4.55  \\ 
$\overline\Xi^+$           &0.0026&0.028 &0.065 &0.145  &0.35  &0.82  \\ 
$\Omega$                   & 0.014&0.065 &0.14  &0.178 & 0.26 &0.39  \\ 
$\overline\Omega$          &0.0008&0.0089&0.014 &0.031 & 0.067&0.16  \\ 
%$\Omega+\overline\Omega$  &0.0145&0.0660&0.131 &0.174 &0.267 &0.399 \\
$\eta $                    & 8.50 &16.7  & 19.5 &23.2  &36.2  &55.6  \\ 
$\eta' $                   & 0.43 &1.13  & 1.06 &1.40  &2.34  &3.78  \\ 
\hline\hline
  $\rho^0 $                  & 11.2 &19.0  & 13.1 &18.9  &30.6  &44.3  \\ 
$\omega(782) $             & 5.94 &12.9  & 11.1 &14.9  &25.7  &38.4  \\ 
$f_0(980)$                 & 0.54 &1.15 & 0.85 &1.21  &2.14  &3.21  \\
${\rm K}^0(892)  $         & 5.72 &12.3 &9.84  &11.9  &17.4  &26.8 \\ 
$\Delta^{0} $              & 37.9 &33.1& 25.16 &26.3  &26.9  &28.0  \\ 
$\Delta^{++} $             & 29.7 &26.08 & 22.18 &24.4  &25.7  &26.7  \\ 
$\Lambda(1520)$            & 1.33 &2.0  & 1.74 &2.11  &2.52  &2.99  \\ 
$\Sigma^-(1385)$           & 2.02 &3.88  & 4.11 &4.51  &5.09  &5.99  \\ 
$\Xi^0(1530) $             & 0.16 & 0.43 & 0.69 &0.84  &1.08  &1.45  \\ 
%$\Theta^+(1540)$           & 0.579&0.953 & 6.74 &6.45  &5.85  &5.94  \\
%$\Xi^{--}(1820)$           &0.0018&0.0086& 0.093&0.11  &0.15  &0.19  \\
\hline
\end{tabular} 
\end{center}
 \end{table}
%%%%%%%%%%%%%%%%%%%%%%%%%%%%%%%%%%%%%%%%%%%%%%%%%%%%%%%%%%%%%%%%%%%%%%%%%%%%%%%%%%%%%

A complete set of particle yields corresponding to the best parameters obtained 
at different reaction energies
is  presented in  table~\ref{outputAGSPS}. Below the stable particle yields,
we show, in bottom section, the resonance yields which are not the measurable
yields, but the initial values required in the study of a 
further resonance evolution~\cite{KUZ}. The stable hadron yields agree
well with the experimental  energy dependence of the yield 
data well, this is, in particular, also the 
case for the $\phi$-yield. However, we expect a growing yield of $\Lambda$, 
which experimentally saturates at 80 $A$GeV and then taking the center of 
measurements point, perhaps even decreases at 158 $A$GeV  compared to 80 $A$GeV.

We believe that imposing strangeness conservation can  
over constrain the fit, a step we only take if the data sample is
small so that  there is no   sensitivity to the $\delta s=0$ 
constraint, as is the case at AGS. 
The values we find for   $\delta s$ at SPS are all negative, see
values  stated in top of table~\ref{outputAGSPS}.
The persistence of $\delta s<0$,
if not result of some experimental NA49 bias, should
be taken as indirect evidence for some new physics. We are
looking forward to RHIC low energy run to resolve this important 
question. Note that the particle yields presented
in table~\ref{outputAGSPS} are evaluated with the value of $\delta s$ as found in fit to 
be the best value. All yields shown in this table are prior to weak decays after hadronization.

%%%%%%%%%%%%%%%%%%%%%%%%%%%%%%%%%%%%%%%%%%%%%%%%%fig 2
\begin{figure}[tb] 
\psfig{width=7.5cm,height=10.8cm,figure=\pathnow 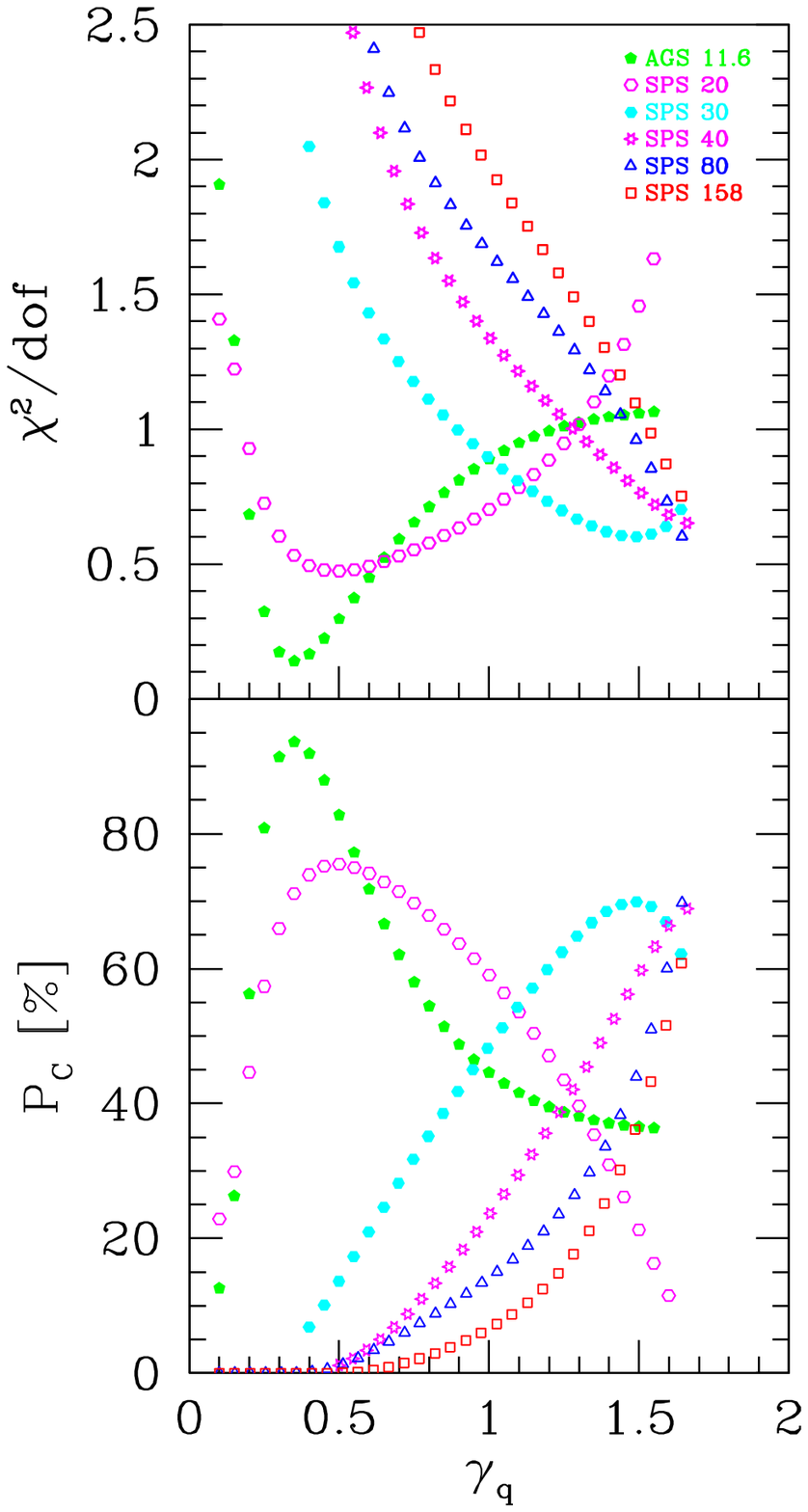} \hspace*{-0.7cm}
\psfig{width=8.5cm,height=10.75cm,figure=\pathnow 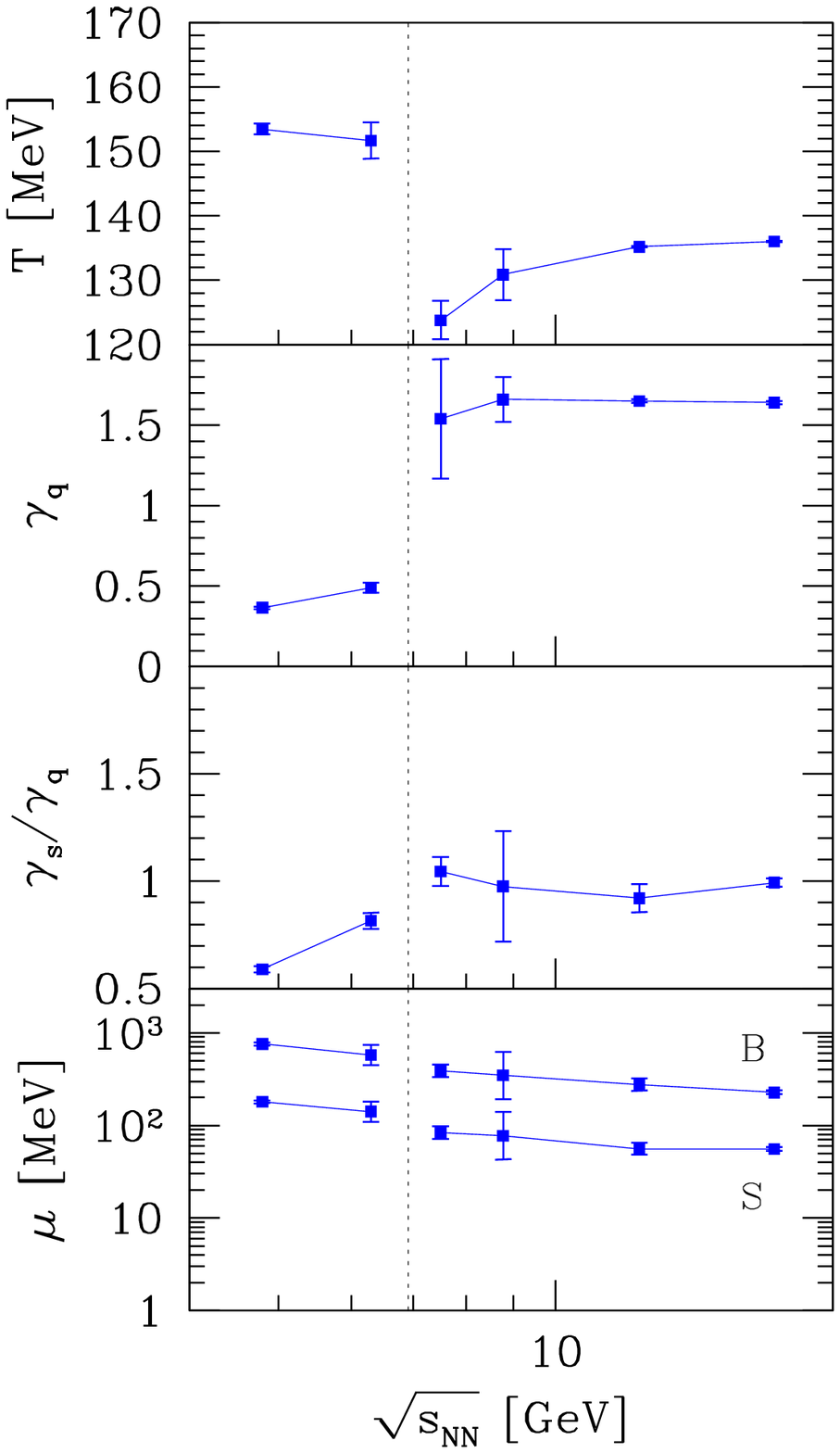}
\caption{\label{ChiP} On left:
$\chi^2\!/{\rm dof}$ (top) and the associated  significance level  $P_{\rm C}[\%]$
(bottom) as function of fixed $\gamma_q$, for the AGS/SPS energy range. On right:
the statistical parameters resulting from a fit at fixed $\gamma_q$, as function of
reaction energy.
}
\end{figure}
%%%%%%%%%%%%%%%%%%%%%%%%%%%%%%%%%%%%%%%%%%%%%%%%%%%%%%%%

We  make, here, an effort to determine if the SH model we use makes good sense
and if the parameters we employ are necessary.  For this purpose,
we show  on left in figure \ref{ChiP}  
in the  top frame   the profile of $\chi^2\!/{\rm dof}$ as 
function of $\gamma_q$, and in  bottom frame, the profile of the 
confidence level  $P_{\rm C}(\chi^2\!,\mathrm{dof})$. We note that there is 
significant variation as function of $\gamma_q$ and that the best fit is not
associated with the value  $\gamma_q=1$. This shows that  this parameter 
not used by many other groups~\cite{PBMequilibrium,Becattini:2005xt,Cleymans:2005xv,Kraus:2007hf}
is neither redundant nor can it assume a fixed 
value, tacitly set to $\gamma_q=1$, 
associated with thermal equilibrium ratio of baryons to mesons.
(note that in the following, we always refer to  $\gamma_q^{\rm HP}$ and thus we sometimes
omit as in this paragraph the superscript).

We see,  on left in  figure \ref{ChiP},  that our fit  
results separate into two groups. The AGS data point and the lowest energy
SPS data point have a clear preference for $\gamma_q^{\rm HP}<1$. This suggests 
that at these two reaction energies (11.6 and 20 $A$ GeV) 
the yield of hadrons and in particular
of  baryons are suppressed, where the benchmark production level is a
chemically equilibrated hadron phase at the temperature $T\simeq 152\pm2$ MeV.
For the 30, 40, 80 and 156 $A$ GeV SPS reactions, we see that our fit results  favor 
enhanced  yields of hadrons and in particular that of 
baryons compared to the benchmark which is a rather 
low chemical freeze-out $T<140$ MeV. 

Sometimes it seems to us  that other groups 
resist the use of $\gamma_q^{\rm HP}$ since they did not incorporate 
this important quantity consistently in their fit programs. Claims that SPS fits which 
include  $\gamma_q$ are unstable were  not true for SPS energy range and 
are certainly false   after the final complete  data set is now
available. Aside of $\gamma_q$, there is another important nuance between 
our and `their' fit. We do not see a value of $\lambda_{I3}$ published by several
other groups, thus we are not sure that this important quantity is employed in the fits,
along with the requirement that net electrical charge of the system is that of the 
participating protons. 

\begin{figure}[tb] 
\psfig{width=8cm,figure=\pathnow 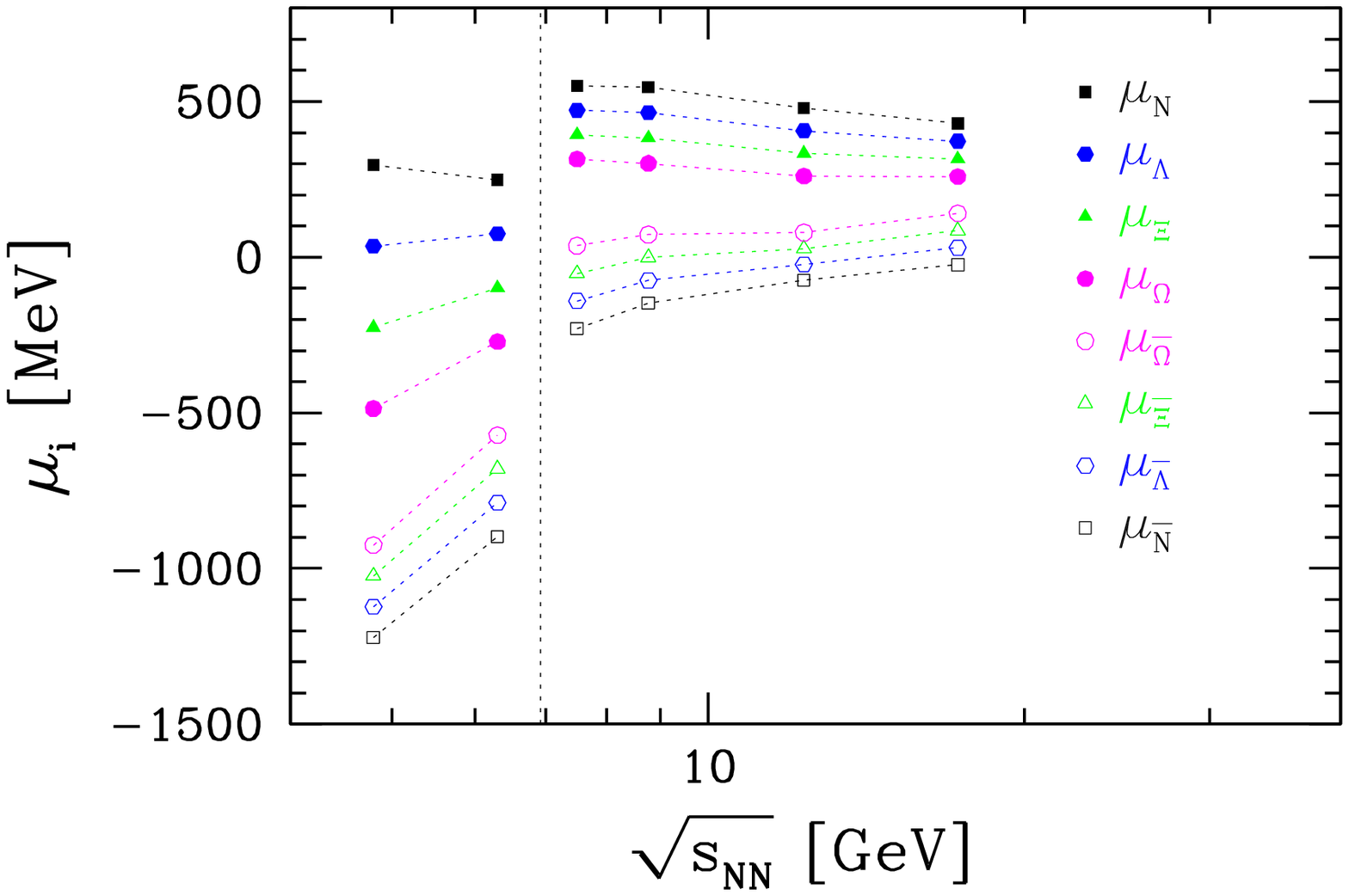} \hspace*{-0.71cm}
\psfig{width=8cm,figure=\pathnow  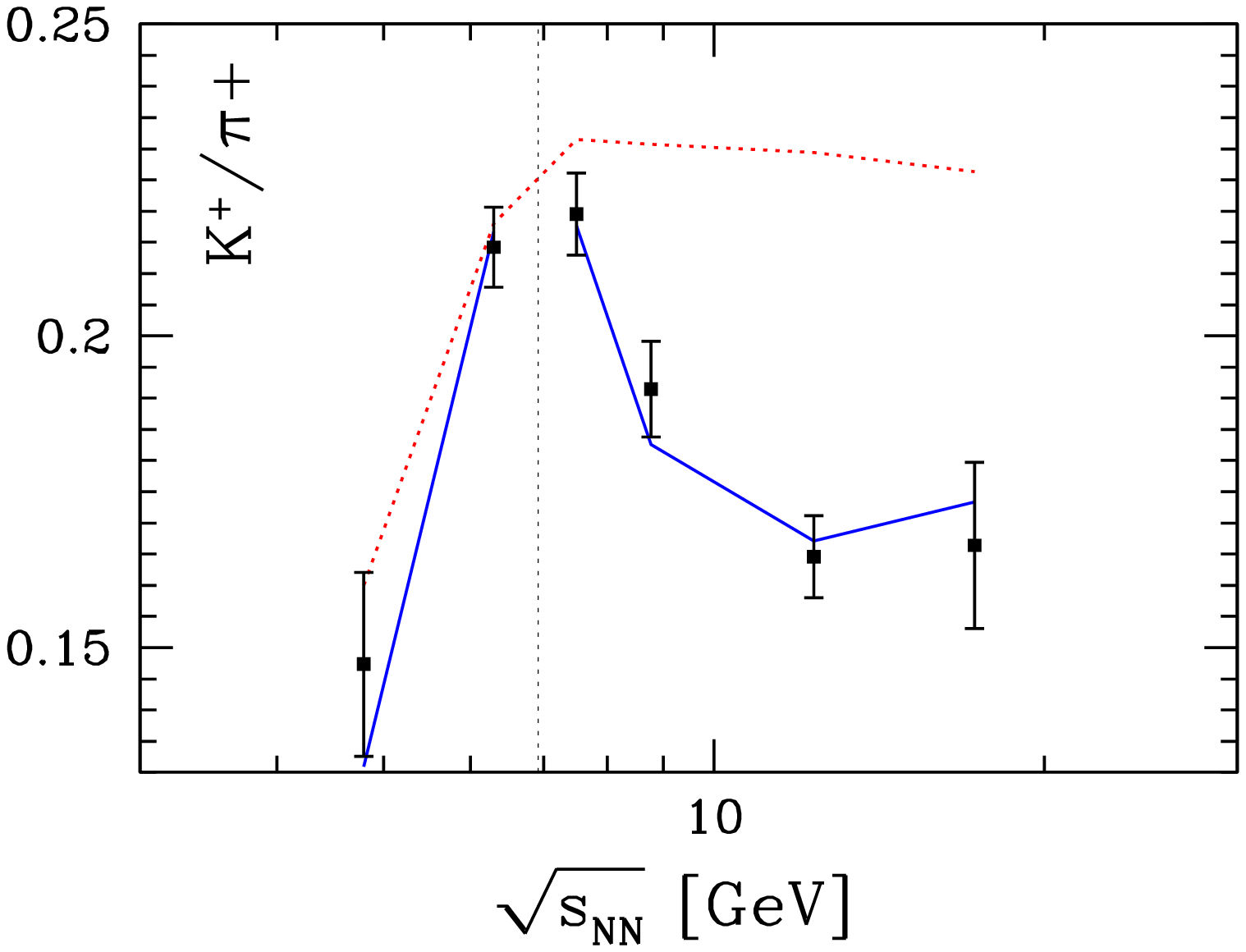}
\caption{\label{PLSQRTSCHEPOTSPSNEW}
On left, the chemical potentials of all baryons and antibaryons in HP. 
On right, our model solid line (blue) tracks K$^+/\pi^+$ yield (data) as 
function of reaction energy. Dotted (red) line:  
equilibrium model~\cite{PBMequilibrium}.
}
\end{figure}

On right in figure \ref{ChiP}, we show the fit results for statistical parameters, 
shown for each $\sqrt{s_{\rm NN}}$. We note that there is a rapid shift in behavior 
 as we go across $\sqrt{s_{\rm NN}}=7$ GeV (vertical dotted line) 
of $T$, $\gamma_q$, and $\gamma_q/\gamma_s$, and of
the volume $V$ seen in the table~\ref{AGSSPS}. This contrasts
with the much  smoother behavior  of the baryo chemical $\mu_B$ and 
strangeness $\mu_S$  potentials (bottom frame on right in figure \ref{ChiP}) 
showing that across the entire reaction energy domain the chemical conditions  
change smoothly. On the other hand, there is considerable shift in chemical 
potentials $\mu_i$  of individual baryons and anti baryons 
at the boundary $\sqrt{s_{\rm NN}}=7$ GeV.
Individual $\mu_i$   are evaluated following the quark content, for example:
$$\mu_{\Xi(ssq)}=2 T\ln (\gamma_s+\lambda_s) + T\ln (\gamma_q+\lambda_q)),\qquad
\mu_{{\overline{\Xi}(\bar s\bar s\bar q})}=2 T\ln (\gamma_s-\lambda_s) + T\ln (\gamma_q-\lambda_q)). 
$$
In figure \ref{PLSQRTSCHEPOTSPSNEW} on left, we see 
that considerable discontinuity arises for 
all baryon chemical potentials between these 
two domains. The energy needed to add a baryon 
is thus smooth, while the energy to add any 
individual particle carrying  baryon number 
is not.
%%%%%%%%%%%%%%%%%%%%%%%%Fig 3
%%%%%%%%%%%%%%%%%%%%%%%%%
 
In figure \ref{PLSQRTSCHEPOTSPSNEW} on right, we compare our fit of K$^+/\pi^+$ with 
data as function of  $\sqrt{s_{\rm NN}}$. While our non-equilibrium model works very 
well and the peak of the horn  we see in the data appears to be  another consequence 
of the change in hadronization behavior near  $\sqrt{s_{\rm NN}}=7$ GeV, the chemical 
equilibrium model~\cite{PBMequilibrium} (dotted, blue)
does not have a priory this capability. We believe 
that in order to explain the NA49 results one needs:\\
a) the strangeness yield parameter $\gamma_s^{\rm HP}$; \\
b) and ability to choose baryon to meson ratio independent of temperature, 
that is  $\gamma_q^{\rm HP}\ne 1$. \\
The above parameters could be de-facto achieved by
 a fine tuned spectrum of hadron resonances \cite{PBMretraction}.
We note that the decreasing value of   K$^+/\pi^+$ beyond the peak
has recently been confirmed by the trial preparatory 
low energy run at RHIC~\cite{Kumar:2008ek}.

In closing this discussion, we note that an analysis of high energy
RHIC data lacks the important constraint to a fixed baryon content, since unlike
at SPS the RHIC results are at central rapidity and the baryon content is one of 
variables which are outcome of the fit. The absence of baryon yield constraint 
reduces the stability of the many-parameter fit and a different approach 
must be taken which we will discuss elsewhere.

%%%%%%%%%%%%%%%%%%%%%%%%%%%%%%%%%%%%%%%%%%%%%%%%%%%%%%%%%%%%%%%
\section{Discussion of data analysis}

\noindent{\bf Q:}  As I see  you do not constrain 
net strangeness to zero --- I think that 
this maybe the reason your effort in this 
field is not taken as seriously as it should!\\ 
{\bf A:} Indeed, we  have gained better 
insight in past 15 years regarding strangeness 
conservation, as compared to our older
 work~\cite{Letessier:1993hi}. At the time we argued that the slowness of the 
weak strangeness decay   assures that in hadronic interactions  {\em net} strangeness
should be zero:  $\Delta s=s-\bar s=0$.  
Today, we allow $\delta s$ to behave like a measurement, that is, we introduce it
with an error $\delta s= 0\pm0.05$. The reasoning is as follows\\
{\bf  a)} This is a test of the hypothesis that weak decays remains weak in QGP phase, and 
the net strangeness remains conserved;\\
{\bf  b)} NA49 did  measure  many, but not all particles carrying strangeness,   e.g.,
$\Sigma^\pm$ has not been measured, this yield 
is uncertain and thus $\delta s= 0$ cannot be
tested to better than about 5\%; \\
{\bf  c)} Summing all measured and unmeasured hadrons  in strangeness
`conservation' condition, 
$\Delta s=\sum_i h_s^i-\sum_jh_{\bar s}^j\to 0$,  combines  independent  measurement
errors and thus, even if NA49 had measured all strangeness carrying hadrons, there would
be a residual statistical  error present in  $\delta s$ --- another way to understand this is 
to note that  we cannot confirm that weak decays in QGP
remain weak to better than the progressing 
error of individual contributing measurements;\\
{\bf d)} Some strangeness could escape detection 
in unknown `particles', for example being bound in 
(nearly) $uds$-quark-symmetric semi-stable strangelett (a small drop of quark matter), this
leads to  $\delta s<0$ --- which is what we find consistently 
in our fits.\\

\noindent{\bf Q:} In what sense is   $P_{\rm C}$ 
(see figure~\ref{ChiP}, left bottom frame) conveying  confidence,
and in `what thing', in what way does this quantity deserve this name, and,
I would think it is better to suppress this part of the figure
as it contains redundant information!\\ 
{\bf A:}   You are   familiar with the Section 32.3 in PDG~\cite{Yao:2006px} 
(PDG: particle data group biannual issue of `Review of particle physics'),
which provides further discussion of the elusive meaning
of $P_{\rm C}$, and figure~32.2  shows  which fixed values of $P_{\rm C}$
arise for  given values $\chi^2\!/{\rm dof}$ and dof. 
The SHARE program evaluates $P_{\rm C}$
for each fit. Abbreviating this section in PDG in a few words: 
When fitting  a statistical data set to a model with 
a few parameters the value of $P_{\rm C}(\chi^2\!,\mathrm{dof})$
predicts (in sense of Bayesian inferment) the 
likelihood of repeat experiment to produce the same outcome 
for the model parameters. Thus,  $P_{\rm C}$  expresses confidence in 
the validity of the model. 

Effectively,  $P_{\rm C}(\chi^2\!,\mathrm{dof})$ 
 also expresses confidence in the data, provided
that we believe in the model:  when we intentionally 
alter an experimental data point  by 2 s.d., our data fit remains
stable in the sense that we find nearly the same model parameters, but $P_{\rm C}$
becomes much smaller, and the one data point which is not well fitted is
the one we altered. Note that  $\chi^2\!/{\rm dof}$ is not as sensitive 
to this consideration, since the effect of one `wrong' measurement is 
diluted by the magnitude  of dof. The function $P_{\rm C}(\chi^2\!,\mathrm{dof})$
turns insensitive   in the limit $\mathrm{dof}\to \infty$, but we are far
from this. In fact, in our study of the fits to the date,
 we found that the value of $P_{\rm C}$ is much more interesting 
when comparing for example NA49 data at different energies since the number of 
available data points varies and thus dof varies along with $\chi^2$. Thus, if we
were to suppress a result it would be the profile of $\chi^2\!/$dof, but that cannot
be done since people want to see  the value of  $\chi^2\!/$dof.

 \noindent{\bf Q:} I note that in your recent work~\cite{Rafelski:2008av}
all 6 best values $P_{\rm C}$ are larger than 0.8 which is very unlikely to happen 
 for 6 independent  measurements.\\ 
{\bf A:}  Yes,  you will further note that in this paper all 
NA49 fits carried out with the final 2008 data set converge 
to a common best value  $P_{\rm C}=70\%$.
This value has higher  $P_{\rm C}$ than one could 
expect on statistical grounds (50\%), but we note
that we have treated systematic errors as if these were statistical errors, and thus
effectively made errors too large. In one case, we also increased the error bar of NA49,
beyond the statistical and systematic error, 
namely we doubled error for $\Lambda$-yield at 80 $A$GeV, 
and thus it is of same relative magnitude as the error at 158 $A$GeV. We did this 
since our fit and $P_{\rm C}$ indicated that this one NA49 value was out of systematics
of other particles measured at this energy. We fit this measurement but with an error
which is modified as stated above.

\noindent{\bf Q:}  How did the SPS heavy ion energy scan  research program come to be?\\
{\bf A:} In early 1980's there were several studies of the mechanism governing the 
development of deconfinement as function of reaction energy. Looking at this work one
finds energy estimates for transition to QGP within a range well beyond the SPS reach. 
Only a  more persistent search will unearth  a minority view, for example 
 in the opening of conclusions of Ref.\cite{Danos:2000mv} we read: 
 ``The formation of a baryon-rich quark-gluon plasma appears 
to be an important reaction channel in collisions of heavy nuclei in the energy 
region of 2.5-5 GeV per nucleon in the center of mass frame of reference''. 
We recall that the K$^+/\pi^+$ peak is at $2\times 3.5$ GeV,
right in the middle of this prediction, obtained considering the growth of  `QGP seeds',
small deconfined drops of matter. Strange particle production  signature of this transformation 
was  the favorite experimental approach~\cite{Rafelski:1996hf}. However, 
only when it was proposed that  an easily  accessible observable K$^+/\pi^+$ could  
suffice~\cite{Gazdzicki:1998vd},   the research program to scan the 
hadron production as function of reaction energy in the SPS domain took off. This 
development is consistent with the  CERN announcement of  February 7, 2000 that 
 a "new state of matter" is produced in  the top SPS energy   central Pb--Pb collisions.

%%%%%%%%%%%%%%%%%%%%%%%%%%%%%%%%%%%%%%%%%%%%%%%%%%%%%%%%%%%%%%%
\section{Physical Properties of the Source}
The SHARE program remains till this day  a unique tool when it comes to the evaluation 
of the physical properties of the source of hadrons. Once a fit is achieved, this is 
a straightforward task, however, with a significant bookkeeping challenge: each hadronic
particle, stable, or resonance, given the statistical parameters $T,\ \mu,\ \ldots$, 
contributes to the physical properties such as pressure $P$, entropy density $S/V$, energy 
density $E/V$, and baryons also to net baryon density  $b/V$, while all strange hadrons 
with s-contents contribute to $s/V$. These properties are of considerable 
importance in discussion  of the meaning of the fit to the data. 

We believe that the extensive properties such as $P$, or $s$  
are  more reliable than the intensive statistical model
parameters. The probability $P_{\rm C}$ is in fact indicating the reliability of 
finding in a redo of experiment and analysis in framework of the same
physical model the same $P,\ S,\ E,\ b,\ s$. This is
so since some elements of our model are not fully established and thus   will evolve in time. 
For example, the hadron spectrum will be in the coming decade  better understood and this 
may alter some statistical properties. However, the occupancy parameters $\gamma_i$ 
compensate to  a large extent just this model dependence, and thus a refinement of hadron 
mass spectrum should  result  in some change of $\gamma_i$, but, as we hope, little changed 
evaluation of the   physical properties of the fireball. Said differently, for example, 
the observed pion yield is a good measure of the entropy produced  with little model dependence.

\begin{figure}[tb] 
\psfig{width=7cm,height=7.5cm,figure=\pathnow 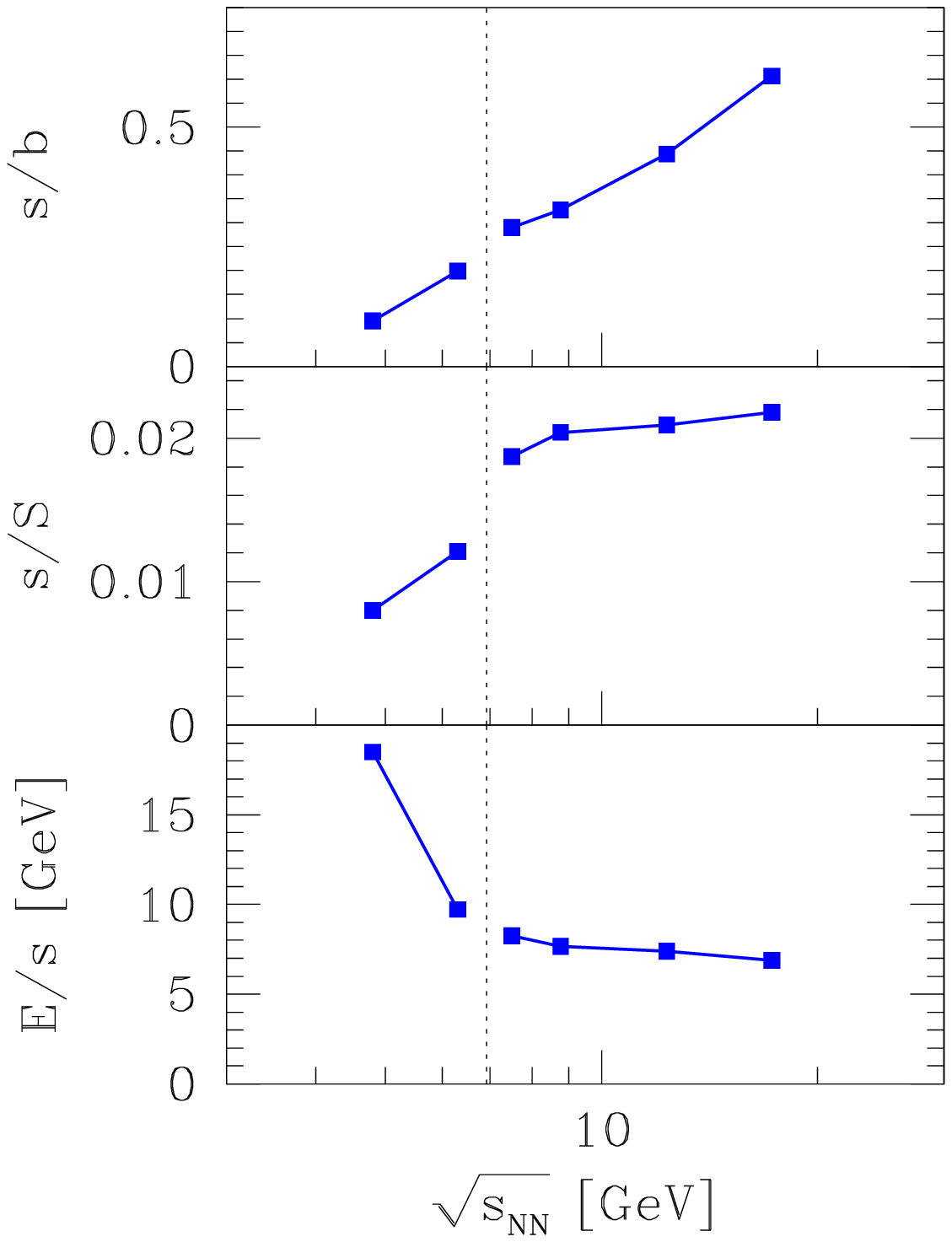}
\psfig{width=7.5cm,height=7.5cm,angle=0,figure=\pathnow 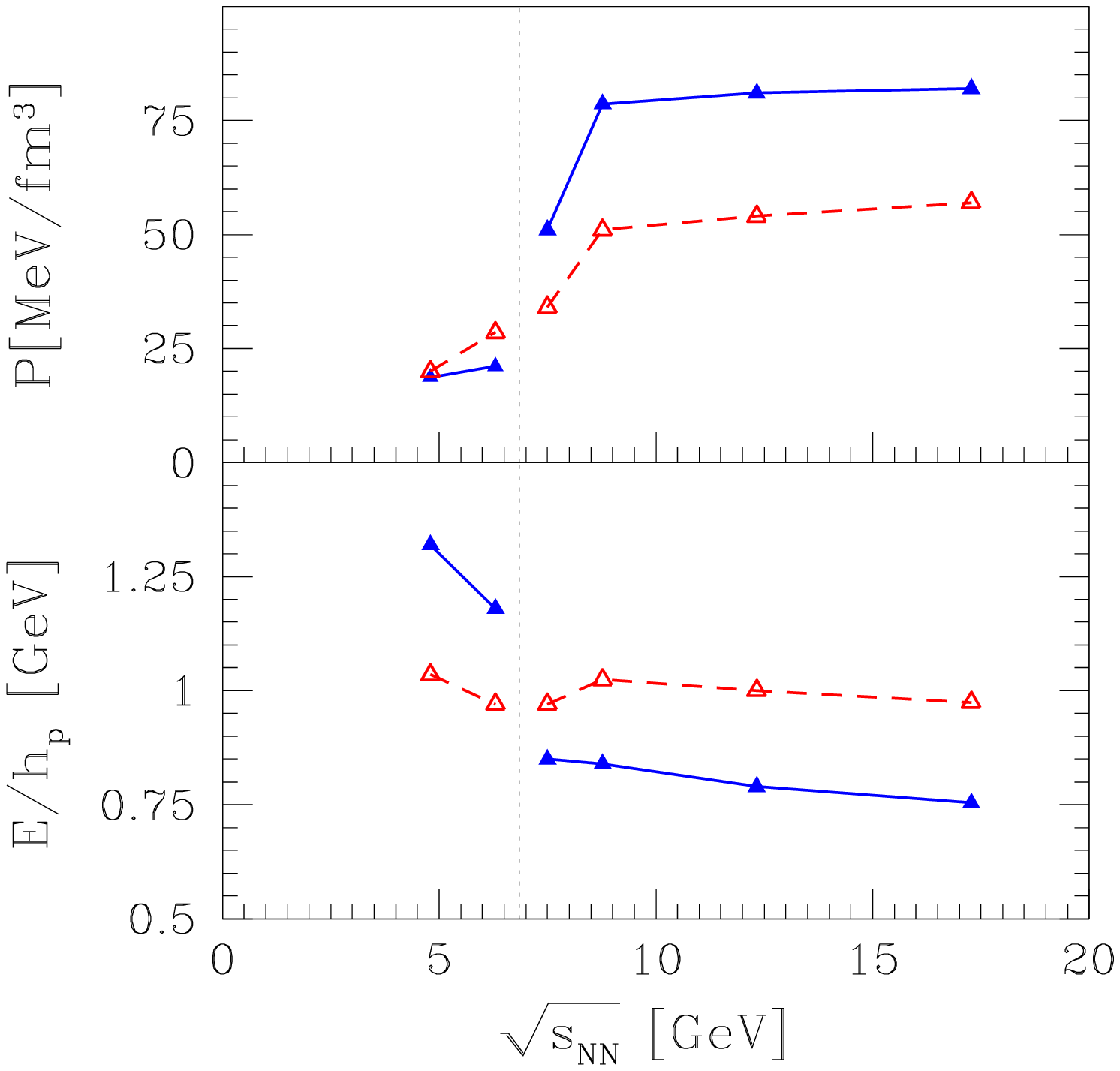}
\caption{\label{PLSQRTSSBSSNEW}
Physical properties of the hadronizing fireball. On left $4s/b,\ s/S,\ E/s$ and 
in right pressure $P$ and thermal energy per primary hadron $E/h_p$. The red
lines on right correspond to $\gamma_q^{\rm HP}=1$ and have a much lower confidence level,
$P_C$, see figure~2.%see figure~\protect{\ref{ChiP}}.
}
\end{figure}

A sample of our findings  for the physical properties 
is presented, in figure~\ref{PLSQRTSSBSSNEW}, as ratio of different
extensive quantities: on left, we see $s/b,\ s/S$, and $E/s$; on right
top frame, the pressure $P$,  and  bottom  right, the thermal energy $E$ per primary hadron $h_{\rm p}$. 
The results we find are the symbols, the
lines guide the eye. All solid (blue) symbols and  lines are full non-equilibrium
model. On right the open, dashed (red) triangles and  lines are semi-equilibrium model with 
$\gamma_q^{\rm HP}=1$. Since $\gamma_q^{\rm HP}$ is artificially fixed, the pressure is 
not the same in both models, and, there is a great reduction of confidence 
level $P_{\rm C}$ in this result, as is seen in figure~\ref{ChiP}, 
left bottom frame:  compare highest 
value of  $P_{\rm C}$ with that arising at a fixed $\gamma_q^{\rm HP}=1$. 
Said differently, the values we present on right in figure~\ref{PLSQRTSSBSSNEW} 
for $\gamma_q^{\rm HP}=1$ (red, dashed) and $\gamma_q^{\rm HP}$ fitted (solid, blue)   
are different since these  are in principle two different 
models to be distinguished by the quality of the fit.

%%%%%%%%%%%%%%%%%%%%%%%%%%%%%%%%%%%%%%%%%%%%%%%Fig4
%%%%%%%%%%%%%%%%%%%%%%%%%%%%%%%%%%%%%%%%%%%%%%%%%%%%%%%%%%%%%%%%%%%%%%%%%%%

Perhaps the most interesting  finding is seen in figure~\ref{PLSQRTSSBSSNEW}
on right in the top frame: the hadronization
pressure is practically constant, near about 82 MeV/fm$^3$ (non equilibrium model, solid line), 
at highest three SPS energies. The absence of fluctuations in this result is most remarkable, 
 a smooth line  is connecting 40, 80 and 158 $A$GeV result. The 
preliminary NA49 data gave less `constant' result~\cite{Rafelski:2008av,Letessier:2005qe}.

The baryon content of the source fireball is fixed 
by the choice of centrality, thus the rise of 
$s/b$ (top left) indicates that strangeness production increases 
rapidly, and  smoothly, with reaction energy. 
Note that  strangeness content increases six fold  
between 10 and 158 $A$GeV  reaction energy,
and by a factor 2.5 between 30 and 158 $A$GeV.
On the other hand,   the   value range $s/S\in (0.019, 0.022)$ 
combined with the growth of $s/b$ shows that 
the fireball  experiences, as function of   reaction energy, a growth in both entropy 
and strangeness at a comparable rate. 

The high value of  $s/S$ 
is established much  earlier in time, in a much more dense phase than the 
 chemical freeze-out. The behavior of $s/S$  
suggests that at 40 $A$GeV (and perhaps event 
at 30 $A$GeV), in an early stage of the fireball
when $s$ and $S$ is made, the same quark--gluon degrees of freedom are already active.
Consideration of kinetic theory~\cite{Alam:1994sc} suggest that  the growth of 
$s/S$ is due to  a gradual as function of reaction energy approach to 
chemical equilibrium, first by gluons, than by light quarks, and later by strange
quarks. We refer to  a more thorough discussion of the importance of  observable $s/S$ 
presented elsewhere~\cite{Letessier:2006wn,Kuznetsova:2006bh}. 

The cost in global thermal energy $E$ 
to make a strange quark pair yield $s$, including those bound 
as pairs in $\eta,\ \eta'$ and $\phi$ is seen 
on left, in the bottom frame of figure~\ref{PLSQRTSSBSSNEW}. 
This value is as low as
6 GeV at highest SPS energy. Here, $E$ is the 
thermal energy content of the hadron 
system and $s$ is as before
the absolute yield of strange quark pairs. 
A strange quark pair at hadronization
has an energy near to 0.9 GeV. The remainder 
of the energy content per this pair, in a nearly 
equilibrated deconfined source of hadrons, is 
contained in the thermal light quarks 
and gluons. This argument accounts well 
for the remaining 5 GeV. The  greater energy required
to produce strangeness pair  at the lower reaction energies is consistent with the 
hypothesis that, prior to hadronization,  strangeness in the fireball
is far below chemical equilibrium, as is seen in the small $s/b$ ratio.

Finally, the bottom frame on right considers the hypothesis that the chemical freeze-out  
condition is related to energy required for making a primary hadron~\cite{Cleymans:2005xv}.
We see that within the semi-equilibrium model (red, dashed line), the value 
is indeed, within a reasonable margin, oscillating near
$E/h_p\simeq 1 $ GeV. However, the high confidence result (blue, solid line)
within the non equilibrium model demonstrates a strong variation 
of $E/h_p$ which drops from above 
1.3 GeV to 0.75 GeV, and thus,  $E/h_p$   is not a good criteria to define freeze-out
of hadrons. Moreover, this variable has a priory no fundamental meaning, and this quantity 
depends strongly on the spectrum of hadrons used, equivalently, the value of $\gamma_q$.

%%%%%%%%%%%%%%%%%%%%%%%%%%%%%%%%%%%%%%%%%%%%%%%%%%%%%%%%%%%%%%%%%%%%
\section{Discussion and Conclusions}
For low reaction energies (AGS 10.6 $A$GeV and SPS 20 $A$GeV), 
we find hadrons originate from a relatively large ($V>3500{\rm fm}^3$), 
dilute ($\gamma_q<0.5$), yet relatively  hot system ($T>150$ MeV). 
The strangeness abundance is low and fast growing. 
At higher reaction energies (SPS 30, 40, 80 and 158  $A$GeV), 
the volume per hadron is less than half as large, 
and the chemical freeze-out proceeds at about 20\%
lower $T$. We know from the experimental results 
that there is a rapid expansion of the 
dense matter. We conclude  that in the  buildup of the matter flow there
was significant cooling of the fireball matter. Particle spectra confirm  that 
there is  fast flow in this SPS energy range. Our  hypothesis, which is consistent
with all results described,  is that we are, 
already at SPS, seeing  sudden hadronization 
of a  supercooled deconfined  phase~\cite{Rafelski:2000by} at 40, 80 
and 158 $A$GeV reaction energy, and to some lesser extend at 30 $A$GeV.   
This hypothesis allows to qualitatively explain the high  $\gamma_q^{\rm HP}>1$ 
value, and is consistent with physical
properties of the system we observe implicitly 
by measuring the hadronization parameters. 

There are several possible explanations of the 
AGS 10.6 $A$GeV and SPS 20 $A$GeV low energy behavior:\\
a) the most straightforward alternative is  the usual 
confined,  hadron phase in which antibaryon annihilation,
which reduces the baryon to meson ratio, takes place, 
which leads to  chemically  under saturated $\gamma_q<1$;\\
b) given the large baryo-chemical potential (see bottom of table\ref{AGSSPS}) 
the freeze-out is in a domain in which a complicated 
QCD phase structure is expected and the physical properties we 
find   are  consistent with 
breakup of a  valance quark deconfined matter~\cite{Letessier:2005qe}. 
Such a phase has relatively  massive constituents 
($m_q\simeq 330,\ m_s\simeq 500$ MeV) 
chiral symmetry is strongly broken, 
even though deconfinement prevails --- only for $\mu_i\to 0$
there is empirical and lattice 
evidence that chiral symmetry and deconfinement are restored at the same condition.
The matter flow of a heavy constituent matter is expected to be slow, and thus, hadronization 
mechanism differs profoundly from what one sees at higher energy;\\
c) Considering relatively high $P_{\rm C}$ near $\gamma_q=1$, for 10.6 $A$GeV we have 45\% and 
for 20 $A$GeV we have 60\%, we could simply discount our results 
at low energies forcing chemical equilibrium  $\gamma_q=1$for 
light quarks, claiming  that   the (small) best 
fit values of $\gamma_q$ are an artifact of   experimental data points. To some extent this 
remark can be made also for 30 $A$GeV data, but cannot be made for 40, 80 and 158 $A$GeV.

Perhaps the most intriguing result of this analysis is the smoothness, and even near 
constancy, of physical properties of the fireball at chemical freeze-out condition
 for the top three  SPS energies 40, 80 and 158  $A$GeV. 
Of particular physical interest is  the value of hadronization pressure  $P\simeq 82$ MeV/fm$^3$,
which is nearly constant. In a phase transformation from quarks to hadrons, the pressure of quarks is 
transferred into the pressure of color-neutral  hadrons, which can escape from the deconfined 
fireball. Since the flow pressure of quarks transfers smoothly into that of hadrons, we conclude that 
the thermal pressure of produced hadrons,  $P\simeq 82$ MeV/fm$^3$, provides a first estimate of 
the pressure of the vacuum which keeps color charged  
quarks inside the fireball up to the point of sudden fireball break-up. 

To conclude, we find analyzing the  final NA49 SPS data 
that  the most central Pb--Pb collisions at  40, 80 and 158  $A$GeV, corresponding to 
$\sqrt{s_{\rm NN}}=  8.76,\ 12.32$, and 17.27  GeV, are showing very similar  behavior at
hadronization, where the physical and statistical 
properties of the source can be interpreted in terms of a 
sudden QGP fireball decomposition. The SPS
reaction at 30 $A$GeV, corresponding to  $\sqrt{s_{\rm NN}}= 7.61$ GeV, 
is an intermediate case, which in many aspects is similar to the higher
 energies, but  the hadronization pressure evaluated in the ensemble of reaction
events is significantly reduced. The reactions 
at  AGS 10.6 and SPS 20  $A$GeV, corresponding to 
$\sqrt{s_{\rm NN}}= 4.84$ and 6.26 GeV, are in a different class 
which cannot be yet fully categorized, and require more precise and comprehensive 
experimental data in order to allow a more specific  conclusion about the applicable reaction 
mechanism. 

\acknowledgments  We thank Marek Ga\'zdzicki for his  interest and valuable comments.
Work supported by a grant from: the U.S. Department of Energy  DE-FG02-04ER4131 and by 
DFG, LMUexcellent.
LPTHE, Univ.\,Paris 6 et 7 is: Unit\'e mixte de Recherche du CNRS, UMR7589

%%%%%%%%%%%%%%%%%%%%%%%%%%%%%%%%%%%%

\end{document}